\documentstyle[aps,prb,multicol,epsf]{revtex}
\begin{document}
\title{Electric field in type II superconductors}
\author{Jan Kol\'a\v{c}ek, Pavel Lipavsk\'y, V\'aclav \v{S}pi\v{c}ka}
\address{Institute of Physics ASCR,
  Cukrovarnick\'a 10, 16253 Prague 6,
  Czech Republic}
\maketitle

\begin{abstract}
Generally it is accepted that electric field ${\bf E}$ in type II 
superconductors is created by the vortex motion, so that it 
is proportional to the vortex velocity ${\bf v}_L$.
This assertion is based on the Josephson relation 
${\bf E} = - {\bf v}_L \times {\bf B}$, which was derived 
and is valid if no transport current is present. 
We present arguments showing that if transport current is present, 
static electric field is proportional to the relative velocity 
of vortices in respect to the velocity of superconducting fluid 
${\bf v}_s$, so that in this case generalised Josephson relation 
${\bf E} = ({\bf v}_s - {\bf v}_L) \times {\bf B}$ is valid. 
\end{abstract}
\begin{multicols}{2}
\section{Introduction}
It is widely believed that electric field ${\bf E}$ in type II
superconductors is generated by the vortex motion, so that it is
proportional to the vortex velocity
\begin{equation} 
{\bf E} = - {\bf v}_L \times {\bf B} ,
\label{J} \end{equation}
where ${\bf B} = n_v \Phi_0$ is the averaged magnetic field, with 
$n_v$ being density of the vortex lattice. 
However, even Josephson in his paper 
\cite{65Josephson} stressed that : 
''The present method is applicable to systems which are 
inhomogeneous with respect to composition or flux line density'', 
so that its applicability is strongly restricted. 
Josephson relation (\ref{J}) can be proved valid also for 
homogenous system in absence of a transport current. 
Here we argue that if the transport current
is present, the averaged electric field is proportional to 
the relative velocity of vortex lattice and superconducting fluid
moving with velocity ${\bf v}_s$,
\begin{equation} 
{\bf E} = ({\bf v}_s - {\bf v}_L) \times {\bf B} . 
\label{gJ} \end{equation}
Before a discussion of physical consequencies of relation (\ref{J}) and
(\ref{gJ}), we want to present a line of simple arguments supporting
(\ref{gJ}):\\
(a) interaction between the superconducting fluid and vortex lattice 
is described by the Magnus force, see e.g.\cite{93Ao},
$
{\bf F}_M(v) = \frac {n_s h}{2}({\bf v}_s - {\bf v}_L) \times {\bf z}
$\\
(b) from the Newton action reaction law it is clear that 
the superconducting fluid must feel the reaction force,
\begin{equation}
{\bf F}_m(s) = -\frac {n_s} {n_v}{\bf F}_M (v)
             = -e({\bf v}_s - {\bf v}_L) \times {\bf B}.
\label{Mr} \end{equation}
(c) in stationary case the reaction force acting on 
superconducting fluid must be compensated by the electric field 
so that $e{\bf E} + {\bf F}_M (s) = 0$ leading to (\ref{gJ}).
\section{The eigenmodes}
Here we will show that the proposed generalisation of the 
Josephson relation is needed in order to ensure that the 
eigenmodes of the system satisfy the momentum conservation law. 
Let us consider the simplest possible case without normal state fluid, 
with only Magnus force acting on the vortex lattice 
(no vortex pinning, no vortex damping, etc). 
The equation of vortex motion is 
\begin{equation} 
{\dot {\bf v}}_{L}  = \Omega ({\bf v}_s - {\bf v}_L) \times {\bf z} , 
\label{v} \end{equation}
where $\Omega = nh/2m_v$ is the angular frequency of 
circular vortex motion with $m_v$ being the vortex mass 
per unit length. Taking into account the third Newton law, 
besides electric field the superconducting fluid must feel 
also the reaction Magnus force (\ref{Mr}), so that its 
equation of motion may be written as
\begin{equation} 
{\dot {\bf v}}_{s}  = e{\bf E} 
         - \omega_c ({\bf v}_s - {\bf v}_L) \times {\bf z} , 
\label{sg} \end{equation}
where $\omega_c = eB/m$ is the cyclotron frequency of 
superconducting charge carriers.
There are two eigenmodes of the system. The zero frequency one
$\omega=0$, $v_L=v_s$ is a direct consequence of the 
Galilei invariance principle - vortices and superconducting fluid may 
move by the same velocity, the total momentum 
$M = n_s m v_s + n_v m_v v_L$ is nonzero, but time 
independent. Using the relation 
$\frac{n_s m}{n_v m_v} = \frac {\Omega}{\omega_c}$  
it is easy to see that the momentum of the  
second eigenmode $\omega = \Omega + \omega_c$ , 
$v_L = -\frac{\Omega}{\omega_c} v_s$ is zero. In this case
vortices and superconducting fluid oscillate with the joint centre 
of mass remaining at rest.
\par
If one would suppose validity of the Josephson relation, 
the equation of motion for the superconducting fluid 
would have to be written as 
\begin{equation} 
{\dot {\bf v}}_{s}  = e{\bf E} 
         + \omega_c {\bf v}_L \times {\bf z} . 
\label{sJ} \end{equation}
The eigen frequencies of the system (\ref{v},\ref{sJ}) are
$\omega_{1,2}= \frac{1}{2}(\Omega 
              \pm \sqrt{\Omega^2 + 4 \omega_c\Omega}$
and the eigen modes are:
$v_L = \frac{1}{2\omega_c} 
     \Omega (\mp \sqrt{\Omega^2 + 4 \omega_c\Omega})$.
According to it the total momentum is nonzero and oscillates 
with the eigen frequency. 
By this the momentum conservation law is violated, 
what clearly shows that equation 
(\ref{sJ}) can not be applied. 
\section{Lorentz transformation}
Let us consider the most simple static case first. 
In laboratory reference frame ($S$) the crystal lattice is at rest
(${\bf v}_c=0$), the averaged magnetic field
inside the superconductor is ${\bf B}$, there is no transport 
current (${\bf v}_s=0$), vortices do not move (${\bf v}_L=0$) 
and consequently electric field ${\bf E}$ inside the 
superconductor is zero. In the reference frame ($S'$)
moving with velocity ${\bf v}$ the vortices and superconducting fluid 
move with the same nonzero velocity ${\bf v}'_L={\bf v}'_s=-{\bf v}$.
From Lorentz transformation it is clear that in this case 
the electric field 
${\bf E}' = {\bf E}+{\bf v} \times{\bf B} = -{\bf v}_L'\times{\bf B}'$
is nonzero and is given by Josephson relation. 
However, for the electrically neutral wire the current density 
is invariant so that it only confirms the fact that without 
transport current the Josephson relation is valid 
in any inertial reference frame even for the homogenous system.

Now let us consider the case that there is transport current 
and vortices move with the same velocity as the averaged velocity 
of the superconducting fluid (${\bf v}_s = {\bf v}_L = -{\bf v}$).
This situation differs from the static case observed from the 
moving reference frame by the fact that crystal lattice 
does not move. 
However, according to the Josephson relation (\ref{J}) electric field
is the same.
If it were true, using Lorentz transformation one finds
that in the reference frame moving with the same 
velocity as the vortex lattice the electric field is zero. 
This consequence contradicts the well known fact that
''Neutral wire with a current appears to be charged 
when set in motion.'' \cite{66Feynman}.
\par
As a result it is clear that if transport current is present, 
Josephson relation can not be applied and must be generalised. 
We should stress that validity of the proposed relation 
${\bf E} = ({\bf v}_s - {\bf v}_L) \times {\bf B}$ 
is restricted to the laboratory reference frame in which velocity 
of the crystal lattice is zero. 
\section{Experimental aspects}
According to generalised Josephson relation the transversal 
electric field is nonzero even in the case that vortices 
are kept by pinning. 
This fact may seem to be in contradiction with transport measurements. 
It is well known, that if vortices do not move, 
the Hall voltage measured with Ohmic contact is 
zero. The probable reason is, that actual Hall voltage 
is canceled by a contact potential. For type I superconductor 
this fact was experimentally proved by Bok and Klein \cite{68Bok} already in
1968.
Type I superconductor measured with Ohmic contacts
also gives zero Hall voltage, but using a contactless 
capacitive pickup to measure changes in the electrostatic potential, 
the Hall voltage, or the so called Bernoulli field 
\cite{68Bok,96Chiang} was be observed. 
We do not know about similar measurement made on 
type II superconductor and this is the way how validity of the 
proposed formula can be tested. 
Far infrared spectroscopy provides experimental data which 
are not influenced by contact potential - and our model 
proved to be consistent with the published experimental results 
(see e.g \cite{99Kolacek}).
In the zero frequency limit the same theoretical approach is
naturally explaining the Hall voltage sign reversal, which is still
considered to be one of the most puzzling phenomena in the physics of
superconductors \cite{98Kolacek}.
%
\section{Conclusion}
If there is no transport current, vortices move only if 
there is gradient of vortex density. 
In this case vortices move and the vortex motion results 
in redistribution of the magnetic flux in and/or outside 
the superconductor. In this case Josephson relation is valid. 
On the other hand, if vortices are moving due to the presence 
of the transport current, vortices are moving, but the averaged 
magnetic flux is time independent. In this case generalized 
Josephson relation must be used. 
%
\acknowledgements
This work was supported by M\v{S}MT program 
''KONTAKT ME 160 (1998)''.


\end{multicols}
\end{document}